\documentclass[%
 reprint,
 amsmath,amssymb,
 aps,
 prl,
]{revtex4-1}

\usepackage{graphicx}
\usepackage{dcolumn}
\usepackage{bm}
\usepackage{natbib}
\usepackage{siunitx}
\usepackage{amsmath}
\usepackage[utf8]{inputenc}

\usepackage{filecontents}

\makeatletter
    \let\@internalcite\cite
    \def\cite{\def\citeauthoryear##1##2{##1, ##2}\@internalcite}
    \def\shortcite{\def\citeauthoryear##1{##2}\@internalcite}
    \def\@biblabel#1{\def\citeauthoryear##1##2{##1, ##2}[#1]\hfill}
\makeatother

\begin{document}

\preprint{APS/PhysRevLett}

\title{Topologically non-trivial photonic surface degeneracy in a photonic metamaterial}

\author{Minkyung Kim}
 \email{kmk961120@postech.ac.kr}
\author{Dasol Lee}%
\author{Dongwoo Lee}%
\author{Junsuk Rho}%
 \email{jsrho@postech.ac.kr}
  \altaffiliation[Also at ]{Department of Chemical Engineering, Pohang University of Science and Technology (POSTECH), Pohang 37673, Republic of Korea}
\affiliation{%
Department of Mechanical Engineering, Pohang University of Science and Technology (POSTECH), Pohang 37673, Republic of Korea
}%

\date{\today}

\begin{abstract}
The recent development of topological photonics has revealed a variety of intriguing phenomena such as Weyl degeneracy. However, topologically non-trivial degeneracy with higher dimension has not been reported in photonics. In this Letter, topologically charged photonic surface degeneracy in a helix structure possessing two-fold screw symmetry is demonstrated. Calculated Chern number and photonic Fermi arcs support topological phase of the surface degeneracy. We also show that the photonic system can work in both metamaterial and photonic crystal regime, which have an order of magnitude different frequency range by the choice of material for the helix. This work suggests the possibilities of unprecedented photonic topological features with higher dimensions.
\begin{description}
\item[PACS numbers]
42.70.Qs, 42.79.Gn, 42.79.Sz
\end{description}
\end{abstract}

\pacs{Valid PACS appear here}
\maketitle

\renewcommand{\vec}[1]{\mathbf{#1}}

Development in topological states have opened fascinating phenomena including quantum Hall effect \cite{klitzing1980new}, quantum spin Hall effect \cite{kane2005quantum}, topologically protected surface states \cite{fu2007topological, zhang2009topological} and Dirac \cite{liu2014discovery} and Weyl \cite{wan2011topological, weng2015weyl, xu2015discovery} degeneracies, and then implemented in classical wave systems such as photonics \cite{PhysRevLett.100.013904, wang2009observation, khanikaev2013photonic, chen2014experimental, wang2016three, guo2017three,yang2017direct, noh2017experimental} and acoustics \cite{he2016acoustic, khanikaev2015topologically, lu2017observation} recently. A Weyl point, which is a topologically non-trivial point degeneracy in three-dimensional momentum space, is a signature of three-dimensional topological phase. Apart from Weyl point which is a zero-dimensional degeneracy, degeneracies have been also found in the higher dimension as nodal line \cite{PhysRevB.92.081201, 1674-1056-25-11-117106, PhysRevB.84.235126} or surface \cite{C6NR00882H, PhysRevB.93.085427} in condensed matter physics. Line degeneracy in a photonic system has been experimentally observed in a metacrystal possessing two orthogonal glide symmetry \cite{gao2018experimental}. However, the line or surface degeneracies in most of the previous researches are generally topologically trivial. Recently, surface degeneracy with non-zero topological charge modeled with tight-binding approximation and its implementation on acoustic metamaterials has been reported \cite{2017arXiv170902363X}. Topological charge of the surface degeneracy relies on the hopping strength, which means that the surface degeneracy can possess no topological charge either.

To best of our knowledge, neither topologically trivial nor non-trivial surface degeneracy has been reported in the realm of photonics. Here, we present a photonic surface degeneracy, which is always topologically charged regardless of geometry parameters as long as the effective parameters are satisfied in contrast to the previous one. The topological surface degeneracy has a partner of double Weyl point located at zero frequency and also support photonic Fermi arcs. The surface degeneracy and its non-zero topological charge are protected by two-fold screw symmetry and effective parameters with hyperbolicity and non-zero chirality respectively. We also demonstrate that our design also has acoustic surface degeneracy, suggesting a possibility of a versatile platform in a variety of classical wave systems covering both photonics and acoustics. Our finding may inspire researches on topologically non-trivial degeneracy in the higher dimension and open new topological features which can be found in classical systems.

\begin{figure} \centering
\includegraphics [width=0.49\textwidth]{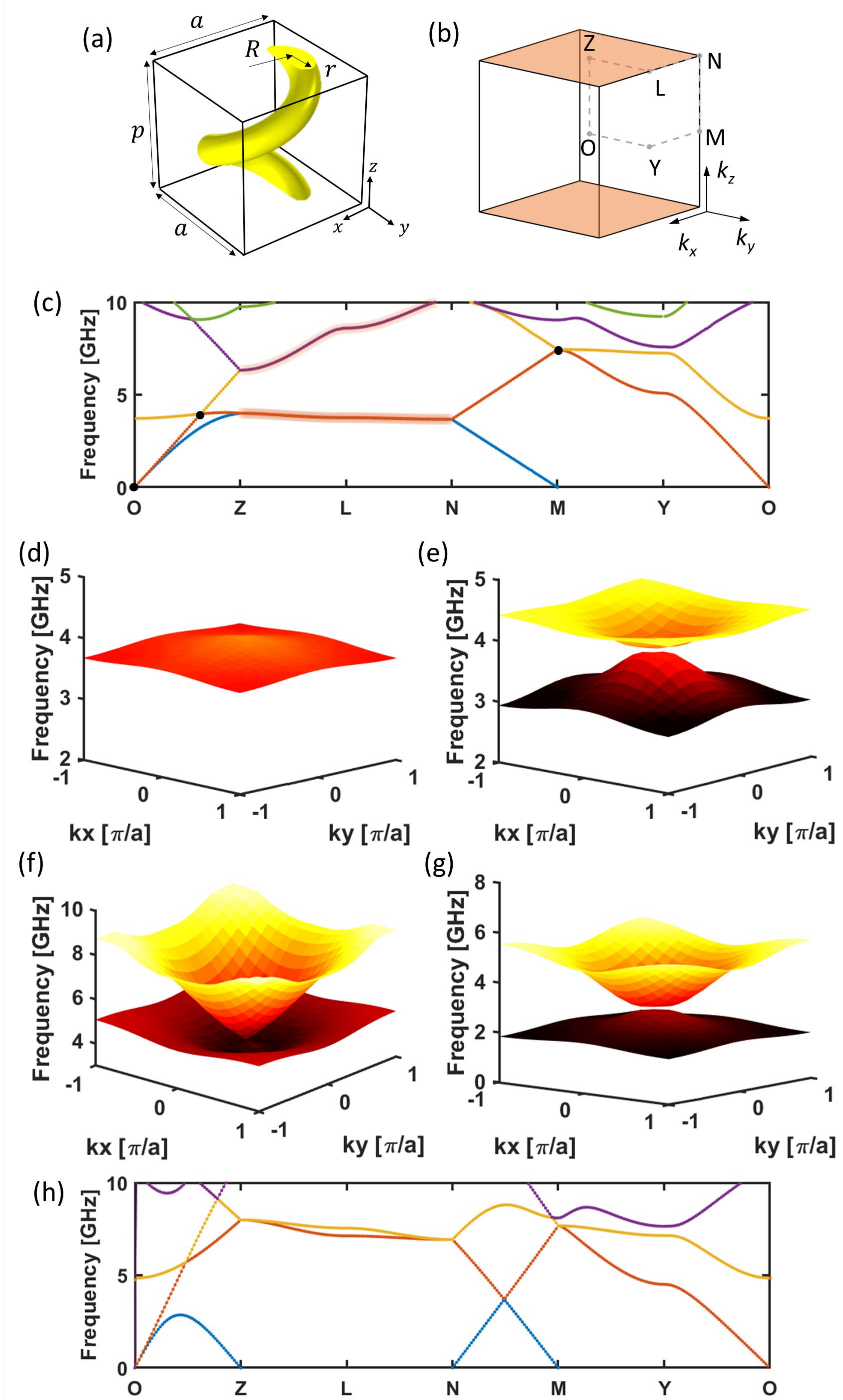}
\caption{(a) Schematic of a unit cell of the helix structure. (b) Brillouin zone with highlighted surface degeneracy (c) Bulk states dispersion along high symmetry lines. Dispersion at (d) $k_z$ = 1 (e) $k_z$ = 0.8 (f) $k_z$ = 0.63 (point degeneracy) (g) $k_z$ = 0.5 (d), (e) and (g) show the first and the second bands while (f) plots the second and the third bands. (h) Bulk states dispersion when two-fold screw symmetry is broken.
}
\label{photonicbulk}
\end{figure}

We employ a periodic array of single helix whose unit cell is shown in Fig. \ref{photonicbulk} as a platform to show topologically charged surface degeneracy. The helix array has been attracted huge attention usually in the field of chiral metamaterials due to the helicity-dependent extinction properties. The helix structure has been also used in a topological system, in which two-dimensional topological phases arisen from the broken inversion symmetry \cite{rechtsman2013photonic}. Here, the helix structure also breaks the inversion symmetry but the mechanism is rather different. The metallic helix provides not only helicity-dependent eigenmodes but also highly anisotropic electric response. In a long wavelength regime, the helix, which is a spatially inhomogeneous system, can be approximated as a homogeneous medium with effective parameters. Then such media can be described by the following constitutive equations.
\begin{align}
    \label{constitutive}
    \vec{D}=\varepsilon_{0}\Bigg[
    \begin{matrix}
    \varepsilon_{x} & & \\
    & \varepsilon_{y} & \\
    & & \varepsilon_{z} \\
    \end{matrix}\Bigg]\vec{E}
    -i\sqrt{\varepsilon_{0}\mu_{0}}\Bigg[
    \begin{matrix}
    \kappa_{x} & & \\
    & \kappa_{y} & \\
    & & \kappa_{z} \\
    \end{matrix}\Bigg]\vec{H}
    \notag\\
    \vec{B}=i\sqrt{\varepsilon_{0}\mu_{0}}\Bigg[
    \begin{matrix}
    \kappa_{x} & & \\
    & \kappa_{y} & \\
    & & \kappa_{z} \\
    \end{matrix}\Bigg]\vec{E}
    +\mu_{0}\Bigg[
    \begin{matrix}
    \mu_{x} & & \\
    & \mu_{y} & \\
    & & \mu_{z} \\
    \end{matrix}\Bigg]\vec{H}
\end{align}
Here, $\varepsilon_{0}$ and $\mu_{0}$ indicate the permittivity and permeability in the free space respectively; $\varepsilon$, $\mu$ and $\kappa$ are permittivity, permeability and chirality, which correspond to electric, magnetic and electromagnetic responses respectively.
The helical structure makes electrons move relatively freely along the helical direction while obstructing along the transverse directions, giving rise to $\varepsilon_{x}=\varepsilon_{y} > 0$ and $\varepsilon_{z} < 0$ when z-axis is defined to be parallel to the helical axis. The opposite sign of effective permittivities is also known as hyperbolicity. Meanwhile, the helical geometry couples electric and magnetic responses and results in non-zero chirality. Combined with hyperbolicity, the chirality enables topologically non-trivial system carrying topologically protected surface states. Topological phases in a photonic system with hyperbolic and chiral properties have been demonstrated theoretically \cite{PhysRevLett.114.037402, xiao2016hyperbolic} and experimentally \cite{yang2017direct}, none of which involves a periodic array of single helix structure. Interestingly, the helix structure supports topologically charged surface degeneracy originating from two-fold screw symmetry described as following:
\begin{equation}
\label{symmetry}
S_{2z}: (x,y,z,t) \rightarrow (-x,-y,z+1/2,t)
\end{equation}
Here, $x$, $y$ and $z$ are normalized by the length of unit cell along each direction. Invariance of the helix structure under the screw symmetry $S_{2z}$ and time reversal symmetry results in band degeneracy in a plane of $k_z = 1$ as an analogy to the Kramers degeneracy in condensed matter physics \cite{PhysRevB.93.085427}.

Bulk states dispersion of the single helix simulated using COMSOL Multiphysics are shown in Fig. \ref{photonicbulk}. We use geometric parameters of $a = 10$ mm, $p = 10$ mm, $R = 3$ mm, $r = 1$ mm. Helical part is modelled as perfect electric conductor while the surrounding medium is set as lossless dielectric whose refractive index $n$ is equal to 2.
Dispersion along high symmetry lines shows the degenerate surface at $k_z=1$, along with several Weyl points marked as circles. Fig. \ref{photonicbulk} (d)-(g) represent dispersion with different $k_z$. Here, (d), (e) and (g) show the two lowest bands. The first and the second bands are degenerate as a surface at the Brillouin zone boundary and then separated as $k_z$ decreases. The surface degeneracies occur not only between the first and the second bands but also between the higher bands. Weyl point between the second and the third band is shown in Fig. \ref{photonicbulk}(f). As the surface degeneracy is protected by two-fold screw symmetry, we deliberately break the symmetry by introducing additional helix structure with half-pitch offset. Then this double helix structure has half periodicity along z-axis, which breaks the two-fold screw symmetry to $S_{z}: (x,y,z,t) \rightarrow (-x,-y,z+1,t)$. Breaking $S_{2z}$ lifts the degenerate surface, generating Weyl points at $k_x = 1, k_y = 1$ instead as shown in Fig. \ref{photonicbulk}.

\begin{figure} \centering
\includegraphics [width=0.49\textwidth]{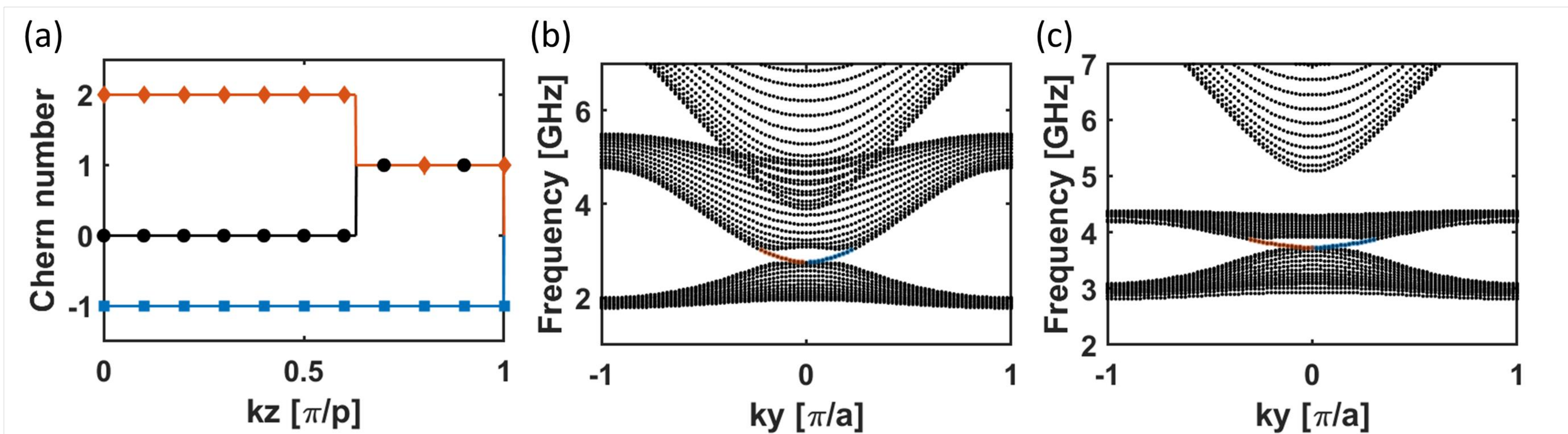}
\caption{(a) Evolution of Chern number along the O-Z line. Lines with square, diamond and circle markers present Chern numbers of the first, the second and the third band respectively. Surface state dispersion at $k_x = 0$ and (b) $k_z$ = 0.5 and (c) $k_z$ = 0.8}
\label{photonicsurface}
\end{figure}

To confirm the topological phase of the degenerate surface, the evolution of the Chern number along O-Z line is shown in Fig. \ref{photonicsurface}(a). We first spatially average the simulated eigenmodes \cite{smith2006homogenization, pendry1999magnetism} and then calculate the Chern number based on Wilson loop method \cite{fukui2005chern}. Chern numbers of the first, the second and the third band are represented as lines with square, diamond and circle markers. Let us now consider degeneracies in each band. The first band has a Weyl point at Brillouin zone center and a surface degeneracy at Brillouin zone boundary. The second band carries additional Weyl points between the third band, one of which is on $k_z$ axis and the other is at point M shown in Fig. \ref{photonicbulk} (c). The third band also shares Weyl points with the upper band. Total Chern number of each three band is neutralized by topologically charged surface degeneracy at zone boundary and Weyl points at zone center.

Topological charge of the surface degeneracy can be confirmed by the distinct Chern numbers between the first and the second band near the zone boundary (Fig. \ref{photonicsurface}(a)). Between topologically inequivalent bands, photonic Fermi arcs exist according to the bulk-edge correspondence. Surface states dispersion at $k_z = 0.5$ and $k_z = 0.8$ are shown in Fig. \ref{photonicsurface} (b) and (c). Since the first and the second band have different Chern numbers, photonic Fermi arcs appear at all $k_z$. On the other hands, Weyl points between the second and the third band equalize the Chern number above $k_z = 0.63$, in which there are no surface states connecting them.
It is worth to note that topological properties in the metallic helix structure appear in metamaterial regime, where the effective medium theory is valid. Within this approximation, choice of lattice system, interaction between neighboring helices or even periodic arrangement does not affect the band dispersion.

\begin{figure} \centering
\includegraphics [width=0.49\textwidth]{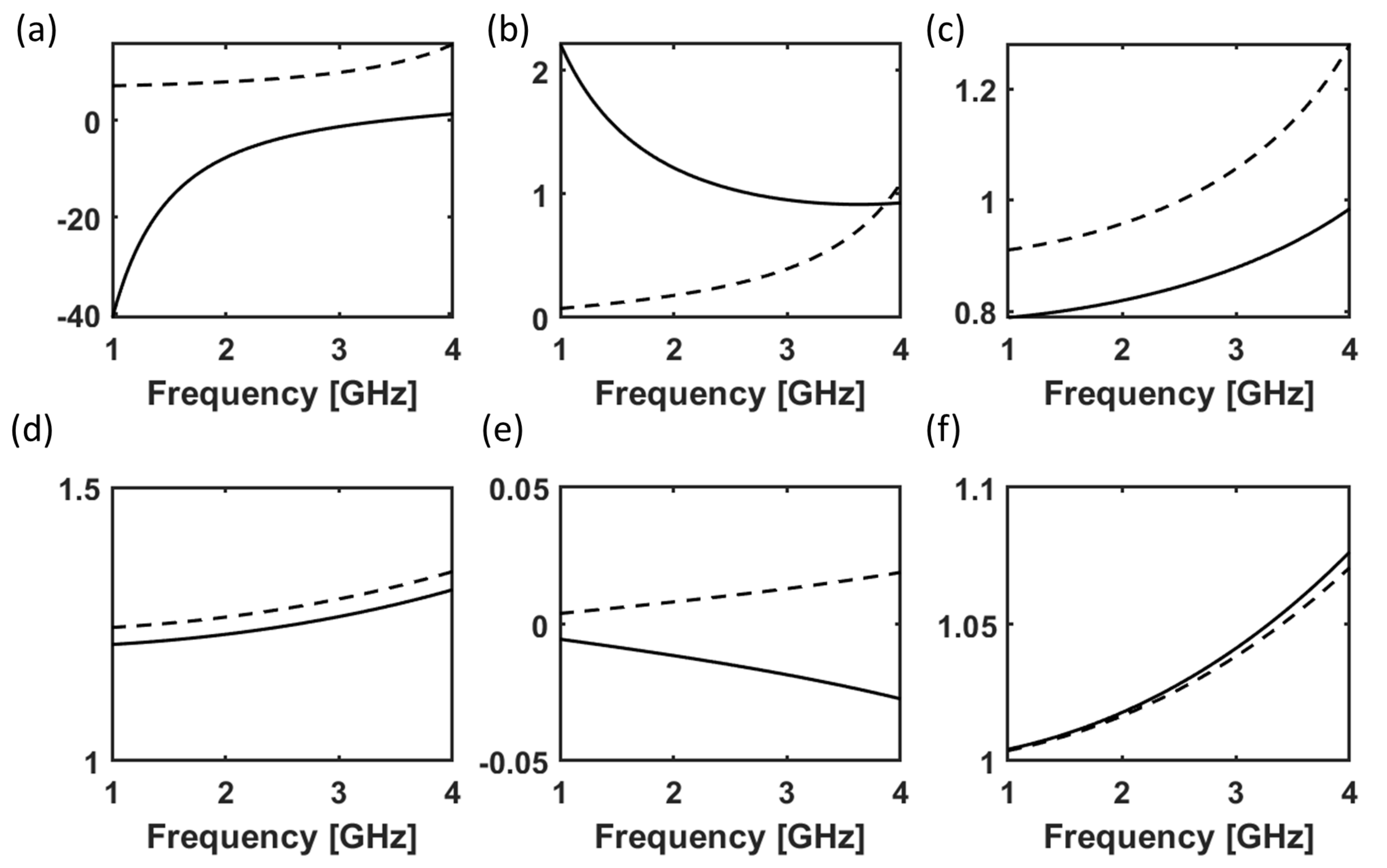}
\caption{Real parts of retrieved effective parameters of (a-c) metallic helix and (d-f) dielectric helix structure. (a) and (d) permittivity, (b) and (e) chirality, (c) and (f) permeability. z and x components are represented as solid and dashed lines respectively. Imaginary parts are skipped here as they are smaller than 0.01 in all cases.}
\label{retrieval}
\end{figure}

The surface degeneracy also occurs for dielectric helix since the surface degeneracy is protected by the combination of two-fold screw symmetry and time reversal symmetry, neither of which is broken here. To examine the difference between metallic and dielectric helix systems, effective parameters retrieved from S-parameters \cite{feng2013effective} are shown in Fig. \ref{retrieval}. Only real parts are plotted since imaginary parts are smaller than 0.01 in all cases. The top and bottom row corresponds to effective parameters of metallic and dielectric helix respectively. Sold and dashed lines indicate component along z- and x-axis. The component associated with y-axis is equivalent to the those with x-axis, and off-diagonal terms are all zeros. Opposite sign of permittivity and non-zero chirality of the metallic helix structure prove hyperbolic and chiral properties. These two properties in low frequency forces double Weyl point at zero frequency, thereby providing non-zero Chern number in the first and the second bands. Since the sum of Chern numbers in each band has to be zero, the surface degeneracy should have also non-zero Chern number to compensate with the zero frequency Weyl point.

\begin{figure} \centering
\includegraphics [width=0.49\textwidth]{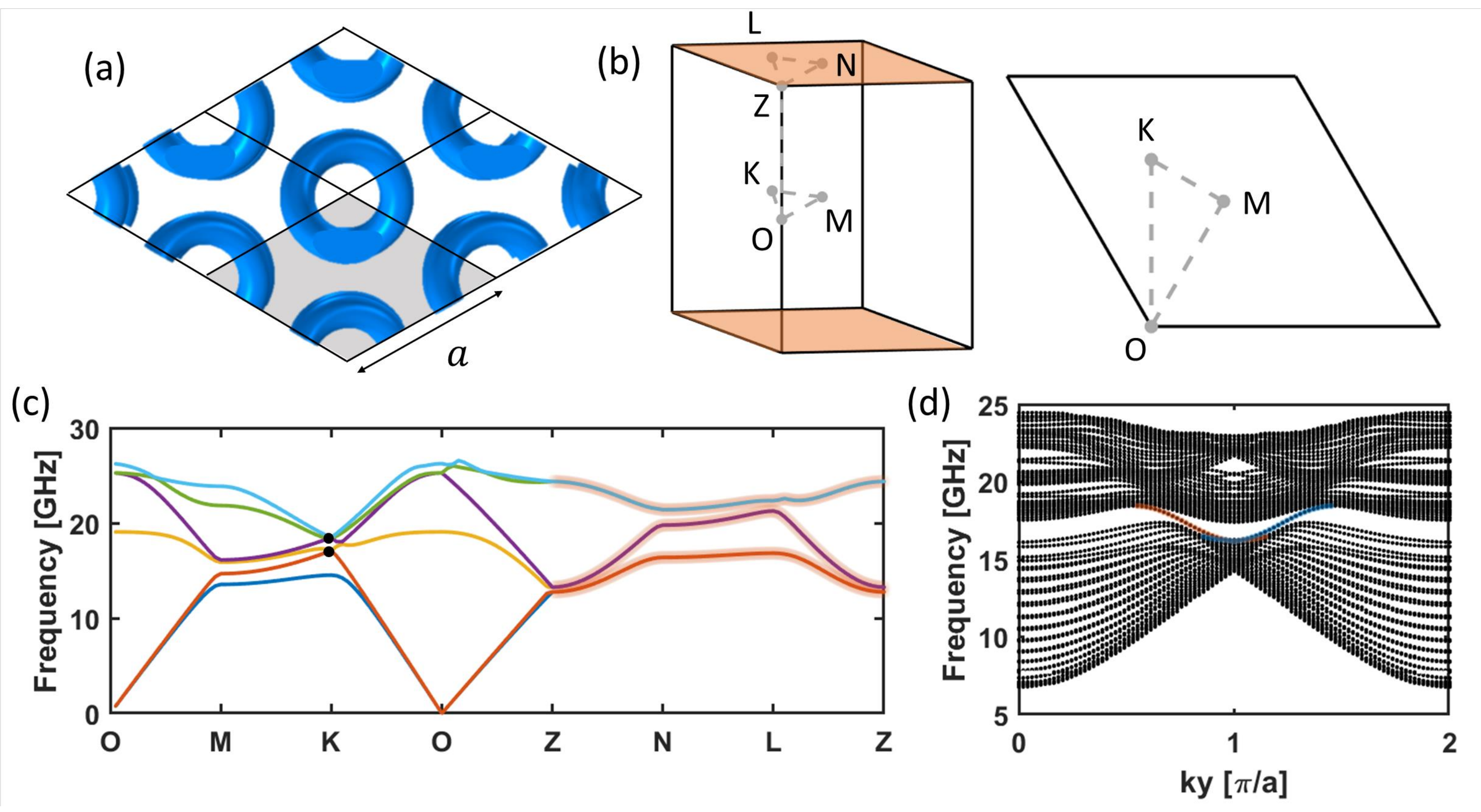}
\caption{(a) Schematic of a unit cell of the dielectric helix structure. (b) Perspective view and the cross-sectional view at $k_z$ = 0 of Brillouin zone with highlighted surface degeneracy (c) Bulk states dispersion along high symmetry lines. (d) Surface state dispersion at $k_z$ = 0.5
}
\label{dielectric}
\end{figure}

In contrast, the dielectric helix has only elliptical dispersion since it cannot support free transportation of electrons along the helix. Furthermore, replacing metal with dielectric weakens the light-matter interaction, leading to negligible chirality. Then topological phase of the dielectric helix cannot be ensured by effective parameters. Despite the loss of hyperbolicity and negligible chirality, the dielectric helix system can also have topological surface degeneracy in photonic crystal regime. Here, topological phase is determined by the type of lattice system and geometric parameters such as lattice constant. Fig. \ref{dielectric} shows bulk and surface states dispersion of the dielectric helix ($n=3$) surrounded by air ($n=1$). We employ a triangular lattice and all geometric parameters are kept the same. As expected, the dielectric helix also supports surface degeneracy at Brillouin zone boundary. The photonic Fermi arcs shown in Fig. \ref{dielectric} (d) prove that the surface degeneracy has non-zero topological charge. The frequency range of the Fermi arcs is an order of magnitude higher than that of metallic helix system, indicating that the dielectric helix system works in photonic crystal regime, not in metamaterial regime.

\begin{figure} \centering
\includegraphics [width=0.49\textwidth]{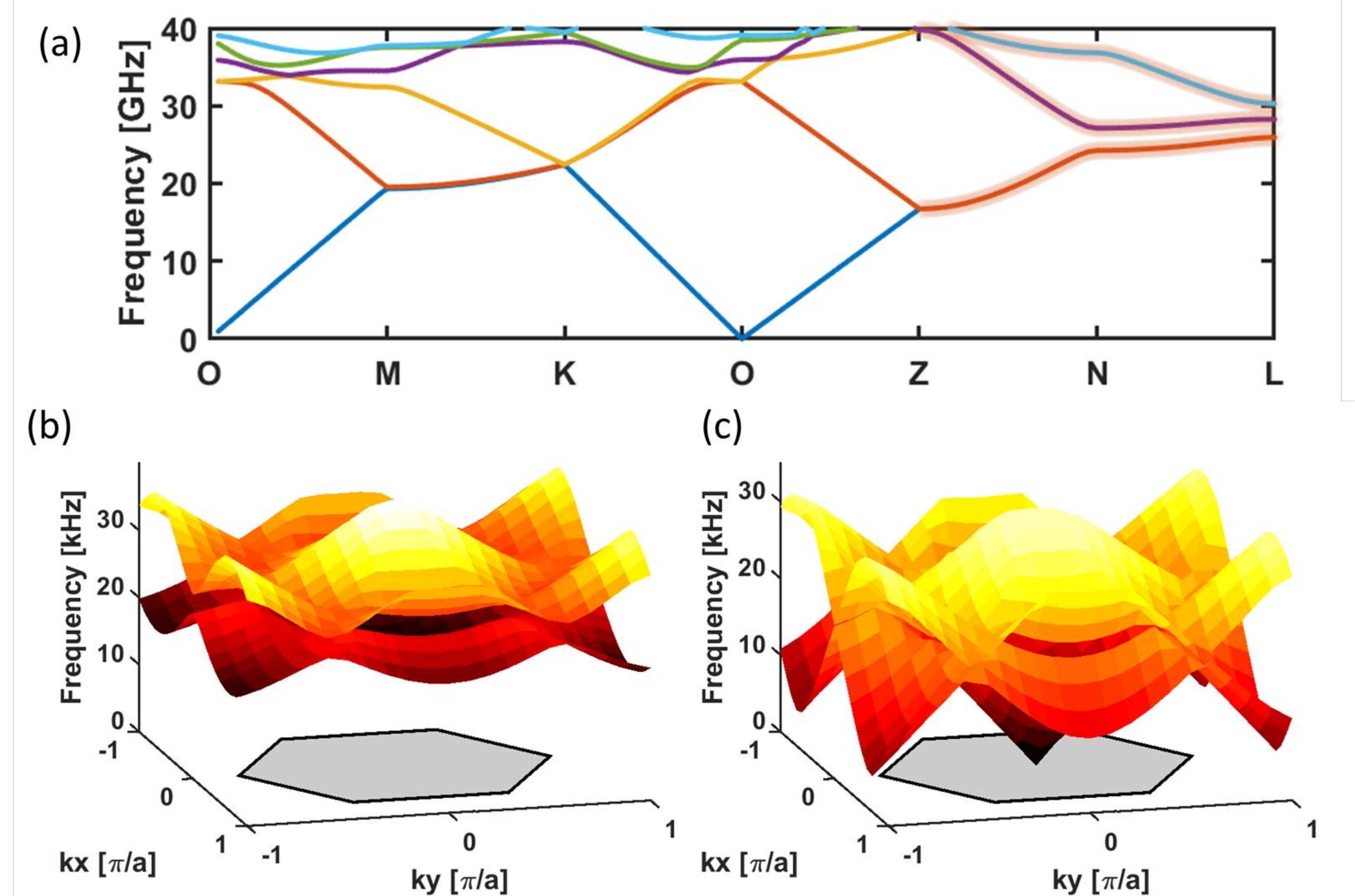}
\caption{Acoustic band diagram (a) Bulk states dispersion along high symmetry lines. Dispersion at (b) $k_z = 1$ and (c) $k_z = 0$. (b) shows the four lowest bands and (c) shows the first and the second bands. The hexagonal pattern on z plane denotes Brillouin zone.}
\label{acoustic}
\end{figure}

As long as the two-fold screw symmetry and time reversal symmetry are conserved, the surface degeneracy should appear in any classical wave systems including acoustics. In this section, we will briefly discuss the possibility of the helix structure in the acoustic system. The helix structure also has acoustic surface degeneracy, where the helix now acts as a rigid material. To calculate acoustic band dispersion, the helix and the background medium are set as lead and air respectively. Note that since most of the solid materials, whether they are dielectric or metal, have density several orders larger than air, the type of materials shows a negligible impact on the band diagram. The helix is arranged in the triangular lattice system equal to that of the dielectric system (Fig. \ref{dielectric} (a)). Geometrical parameters are kept same as the previous one. Fig. \ref{acoustic} (b) shows the four lowest bands at $k_z=1$ while (c) shows the first and the second bands at $k_z=0$. At the Brillouin zone boundary, 1st-2nd and 3rd-4th band are being degenerate, generating only two distinct bands. Although the degenerate surfaces are not topologically charged in this case, engineering the band gap by controlling system parameters such as lattice type and lattice constant will lead to topologically non-trivial surface degeneracy in acoustics as shown in \cite{2017arXiv170902363X}.

In conclusion, we present topologically non-trivial surface degeneracy in a helix structure. In metallic helix structure, topological phase of the surface degeneracy is protected by effective properties of the structure in a long wavelength regime. Therefore, designing such a system is free from choosing adequate lattice system and precise control of geometric parameters as long as chiral and hyperbolic properties are guaranteed. On the other hand, dielectric helix structure in photonics systems, which does not satisfy the effective parameter conditions, may have topological surface degeneracy in photonic crystal regime. It still possesses the surface degeneracy but its topological charge is non-zero only when the system is well-designed. We demonstrate that simply replacing metal by dielectric gives rise to topologically non-trivial surface degeneracy at frequency one order of magnitude higher. The helix structure also supports surface degeneracy in acoustics, in which the helical parts play a role of rigid materials to locally block the propagation of acoustic waves. We believe that the existence of topologically charged surface degeneracy implies a variety of unrevealed topological features in not only photonics but also in other classical wave systems.

\vspace{2mm}
This work was financially supported by the National Research Foundation of Korea (NRF) grants (NRF-2018M3D1A1058998, NRF-2015R1A5A1037668 and CAMM-2014M3A6B3063708) funded by the Ministry of Science and ICT (MSIT) of the Korean government. M.K. and D.L. acknowledge Global Ph.D. Fellowships (NRF-2017H1A2A1043204 and NRF-2018H1A2A1062053, respectively) from NRF-MSIT of the Korean
government.

\bibliographystyle{apsrev4-1}
\bibliography{\jobname}

\end{document}